\documentclass{cernyrep}

\usepackage{endnotes}       
\usepackage{color}
\def\rmi{\mathrm{i}} 
\def\rmd{\mathrm{d}} 
\def\rme{\mathrm{e}} 
\def\usp{\:}         
\usepackage{amsfonts}
 \DeclareSymbolFont{upgreek}{U}{eur}{m}{n}
 \DeclareMathSymbol{\rmalpha}{0}{upgreek}{"0B}
 \DeclareMathSymbol{\rmbeta}{0}{upgreek}{"0C}
 \DeclareMathSymbol{\rmgamma}{0}{upgreek}{"0D}
 \DeclareMathSymbol{\rmdelta}{0}{upgreek}{"0E}
 \DeclareMathSymbol{\rmmu}{0}{upgreek}{"16}
 \DeclareMathSymbol{\rmpi}{0}{upgreek}{"19}
 \DeclareMathSymbol{\rmsigma}{0}{upgreek}{"1B}
 \DeclareMathSymbol{\rmphi}{0}{upgreek}{"1E}
 \DeclareMathSymbol{\rmomega}{0}{upgreek}{"21}

\usepackage{texnames}
\usepackage[T1]{fontenc}
\begin{document}
\title{Direction for the Future -- Successive Acceleration of Positive and Negative Ions Applied to Space Propulsion}

\author{A. Aanesland, J. Bredin, L. Popelier and P. Chabert}%

\institute{Laboratoire de Physique des Plasmas, CNRS -- Ecole Polytechnique, France}

\maketitle 

\begin{abstract}
Electrical space thrusters show important advantages for applications in outer space compared to chemical thrusters, as they allow a longer mission lifetime with lower weight and propellant consumption. Mature technologies on the market today accelerate positive ions to generate thrust. The ion beam is neutralized by electrons downstream, and this need for an additional neutralization system has some drawbacks related to stability, lifetime and total weight and power consumption.
Many new concepts, to get rid of the neutralizer, have been proposed, and the PEGASES ion--ion thruster is one of them. This new thruster concept aims at accelerating both positive and negative ions to generate thrust, such that  additional neutralization is redundant.  This chapter gives an overview of the concept of electric propulsion and the state of the development of this new ion--ion thruster.
\end{abstract}

\section{Introduction}

In the late 1950s, the golden years of space exploration began. The technology improvements in rocketry after World War II allowed us to meet the challenge of creating thrust that could overcome Earth's gravity. Human exploration to the Moon became a reality; and also the huge technology breakthrough led to increasing numbers of satellites in orbit around the Earth and unmanned missions to our neighbouring planets, comets, asteroids, etc. The early \textit{Pioneer} and \textit{Voyager} missions have now reached interstellar space and still transmit signals back to Earth.

Electrical space thrusters show important advantages for applications in outer space compared to chemical thrusters, as they allow a longer mission lifetime with lower weight and propellant consumption. The first mission that used electric propulsion as a unique propulsion system was launched in 1989 with the \textit{Deep Space~I} programme. Since then, this technology has become more and more popular within the space sector, but still this makes up only a few per cent of the launched missions.
The  mature technologies on the market today rely on accelerating ions either via the Hall current (i.e., Hall thrusters or closed drift thrusters) or via electrostatic grids (i.e., ion thrusters) \cite{kaufman:1985aj0,goebel:20080}.
Hall thrusters and gridded thrusters have various advantages and drawbacks, and it is mainly political issues that determine which one is used on a mission. The advantages of gridded thrusters (over Hall thrusters) are lower beam divergence, slightly longer lifetime and the fact that they can deliver higher thrust and ion beam velocity.
The common point is that positive ions are accelerated from the plasma via electric fields to generate thrust. As the beam is positively charged, electrons are injected into the downstream space to ensure current and charge neutralization.
This need for downstream neutralization is a drawback in all existing systems, as it takes up space and adds to the weight of the system. It also increases the complexity of the thrusters and consequently the risk of failure. The \textit{HAYABUSA} mission is one example illustrating this problem \cite{hayabusa}.

The PEGASES thruster concept has been proposed to improve electric propulsion and to remove the need for additional neutralization systems \cite{Aanesland:2009jm}. PEGASES is an acronym for `plasma propulsion with electronegative gases'. It belongs to the electrostatic gridded thruster family, but, contrary to classical systems, it accelerates alternately positive and negative ions to provide thrust. In this way additional electron neutralization is redundant. It is also thought that, since the recombination rate of oppositely charged ions is much higher than that for electron--ion recombination, the downstream beam or plume will mainly consist of fast neutrals and a low density of charged particles.

This chapter will outline the various stages of the PEGASES thruster, from plasma generation to the formation of an electronegative plasma and the acceleration of both positive and negative ions to provide thrust.

\section{Thrust and specific impulse in space propulsion}

The thrust $T$ acting on a spacecraft  is given by the change in momentum such that
\begin{equation}
T=-\frac{\rmd (mv)}{\rmd t}=-v_{\rm ex}\frac{\rmd m}{\rmd t},
\label{eq:momentum}
\end{equation}
where $v_{\rm ex}$ is the exhaust velocity and $\rmd m/\rmd t$ is the rate of change of the total mass ejected.
The thrust required in space applications can vary from meganewtons to micronewtons. To escape the Earth's gravitational field, the thrust needs to overcome the weight of the spacecraft. This requires very high thrust, which today can only be reached by chemical rockets that expel a large amount of mass very quickly (i.e., a large $\rmd m/\rmd t$). However, when in space, the thrust can be much lower. Sometimes, only micro-thrust is needed to precisely position space vehicles with respect to one another, and at other times the goal is to reach a target with as little propellant as possible but with no strict requirement on the mission duration. With low thrust, the spacecraft will not travel as fast, but will eventually reach the target.

\begin{figure}[htbp]
\centering\includegraphics[width=0.5\linewidth]{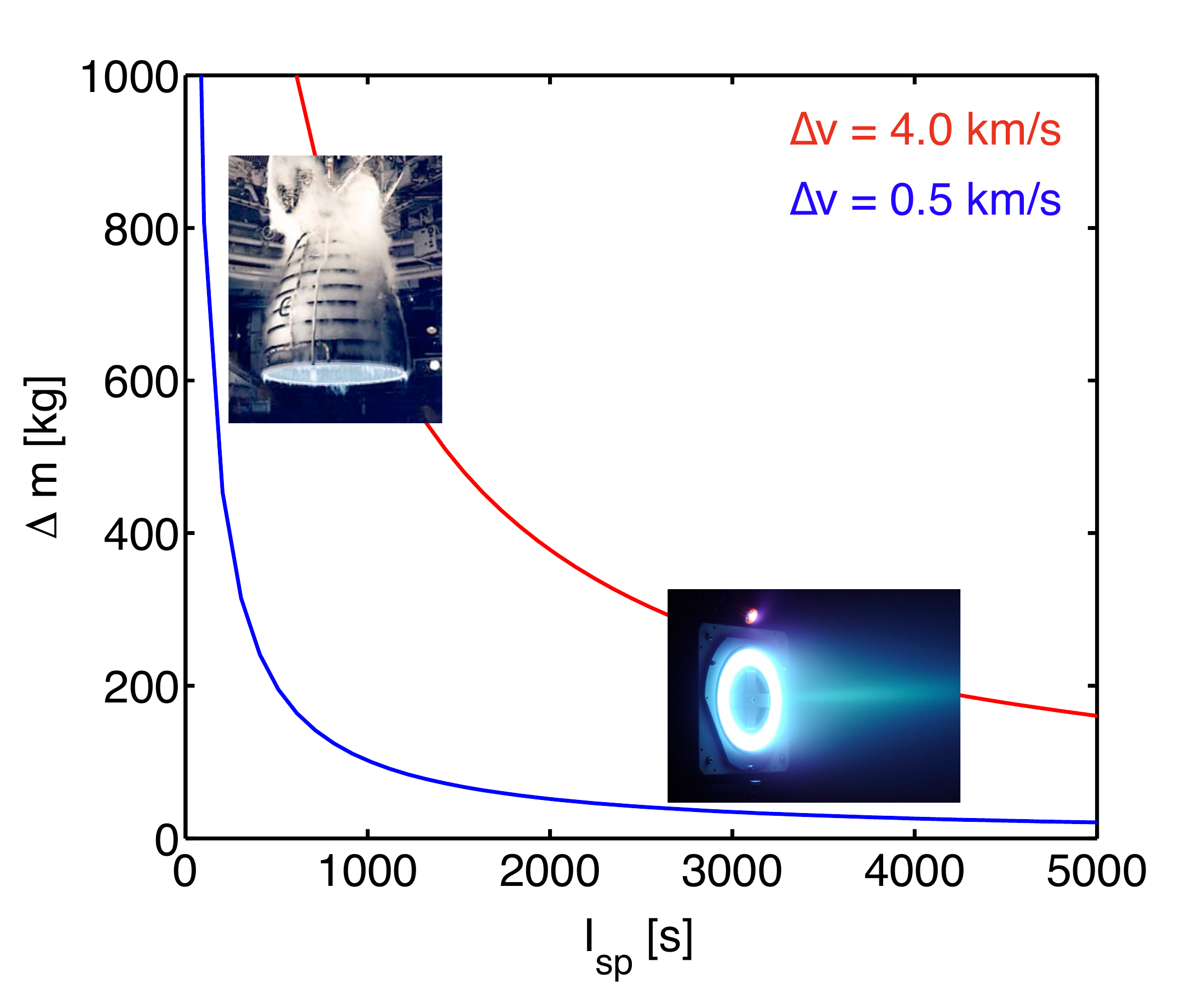}
\caption{Propellant mass required as a function of the specific impulse for two different types of missions}
\label{fig:mass_vs_Isp}
\end{figure}

Integrating Eq.\ \eqref{eq:momentum} gives the well-known rocket equation,
\begin{equation}
\Delta v = v_{\rm ex} \ln\left(\frac{m_0}{m_{\rm f}}\right),
\label{eq:delta_v}
\end{equation}
where $\Delta v$ is the change in velocity of the space vessel, and $m_0$ and $m_{\rm f}$ are its initial and final mass, respectively. Hence, $\Delta m=m_0-m_{\rm f}$ is the required propellant mass.
This equation was first derived by Konstantine E. Tsiolkovsky (1857--1935) and is sometimes referred to as the Tsiolkovsky rocket equation.
The specific impulse is defined as
\begin{equation}
I_{\rm sp}=\frac{v_{\rm ex}}{g_0},
\end{equation}
and is given in the unit of seconds ($g_0$ is the gravitational constant at sea level).
Eq.\ \eqref{eq:delta_v} shows that the $I_{\rm sp}$ of a thruster is indirectly a measure of the propellant consumption: the higher $I_{\rm sp}$, the less propellant is needed to reach the same change in velocity. Figure\ \ref{fig:mass_vs_Isp} is an example given for an interplanetary mission that will need a velocity change $\Delta v$ of 0.5{\usp}km{\usp}s$^{-1}$, or for a positioning in low Earth orbit with a velocity change of 4{\usp}km{\usp}s$^{-1}$. In the first example, to achieve the same manoeuvre one could go from a propellant mass of around 400--800{\usp}kg to only 25{\usp}kg by choosing electric rather than chemical propulsion. Yet, the duration to achieve the manoeuvre is very different due to the thrust level. To bring 1{\usp}kg up to low Earth orbit with an Ariane~V rocket costs around 20\,000 euros \cite{Janovsky}. Hence reducing the propellant consumption translates into real money for the industry. In the example above, a mission to the Moon saves 7--15 million euros.

\subsection{Thrust from gridded ion thrusters}

For an ion beam, $v_{\rm ex}$ is equivalent to the ion beam velocity $v_{\rm b}$ and $\rmd m/\rmd t$ can be expressed as the ion flux $\Gamma_\rmi$ out of the thruster with the given mass $M_\rmi$ through an effective grid area for ions $A_\rmi$. Assuming that $v_{\rm b}=\sqrt{2eV_0/M_\rmi}$ is obtained by accelerating the ions across a voltage difference $V_0$, we can express the thrust provided by an ion beam as
\begin{equation}
T_\rmi=A_\rmi \Gamma_\rmi M_\rmi v_{\rm b} = A_\rmi e n_{\rm s}\sqrt{2T_\rme V_0}.
\end{equation}
Hence, the thrust depends on the plasma density at the sheath edge $n_{\rm s}$, the electron temperature $T_\rme$, the acceleration voltage $V_0$, and the area of active extraction/acceleration $A_\rmi$.
When the thruster is operated at its maximum performance, the current is limited by the space-charge-limited current through the grids, given by the Child--Langmuir relation \cite{kaufman:1982}
\begin{equation}
J_{\rm CL}= \frac{4}{9} \varepsilon_0 \left(\frac{2e}{M_\rmi}\right)^{1/2}\frac{V_0^{3/2}}{d^{*\,2}},
\label{eq:J_CL}
\end{equation}
where $d^*$ is the effective space-charge-limited distance between the grids. Assuming that the plasma density and hence the flux from the plasma is balanced with the space-charge-limited current, such that
$\Gamma_\rmi=J_{\rm CL}/e$, the maximum thrust can be expressed as
\begin{equation}
T_\mathrm{max}=\frac{8}{9}\varepsilon_0 \frac{A_\rmi}{d^{*\,2}}V_0^2.
\end{equation}
Here we neglect beam divergence, which would reduce the thrust by $\cos^2\theta$, where $\theta$ is the beam divergence angle, and we neglect any losses in the grid system.
Hence, provided that the plasma is of sufficient density, the thrust depends only on the grid dimensions and the acceleration voltage. Higher thrust is achieved with larger surfaces and smaller grid distance (or, to be precise, a smaller  space-charge-limited distance).
As a short remark, note that the mass of the ions does not effect the thrust under these conditions. However, the mass plays a significant role in the specific impulse, and hence the propellant consumption.

\begin{figure}[htbp]
\centering\includegraphics[width=0.98\linewidth]{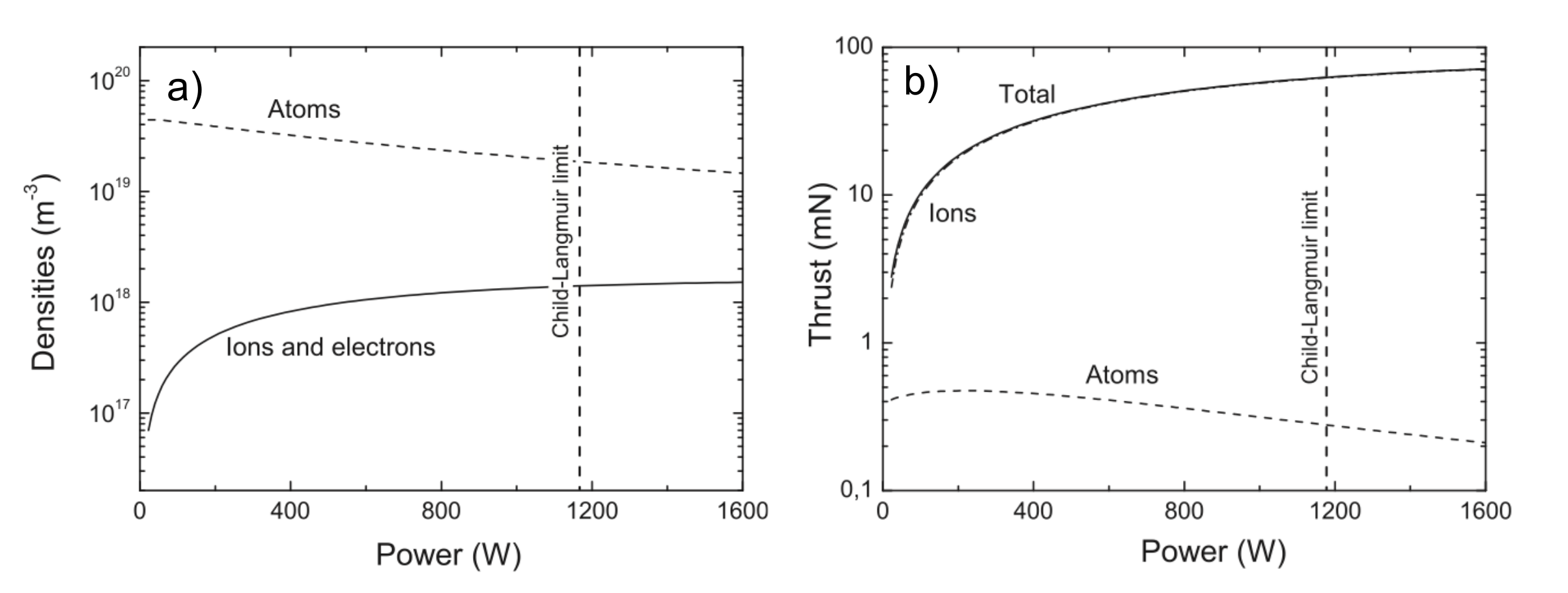}
\caption{(a) Atom and ion densities and (b) corresponding thrust as a function of power. The dotted line indicates the Child--Langmuir space-charge limit for the grids used in this calculation ($d=1${\usp}mm).}
\label{fig:model-pascal}
\end{figure}

A global (volume-averaged) model of a gridded ion thruster has recently been developed \cite{Chabert:2012dh}.
The neutral propellant (xenon gas) is injected into the thruster chamber at a fixed rate and a plasma is generated by circulating a radio-frequency (RF) current in an inductive coil. The ions generated in this plasma are accelerated out of the thruster by a pair of d.c. biased grids. The neutralization downstream is not treated. Xenon atoms also flow out of the thruster across the grids. The model, based on particle and energy balance equations, solves for four global variables in the thruster chamber: the plasma density, the electron temperature, the neutral gas (atom) density, and the neutral gas temperature.
Figure\ \ref{fig:model-pascal} shows (a)~the densities and (b)~the thrust as functions of the input RF power, for a gridded system where the grids are separated by 1{\usp}mm.
Intuitively, the thrust increases with increasing plasma density, as the current or flux from the plasma increases. What is less intuitive is that the maximum thrust, limited by the space charge, is reached for rather low RF power and plasma density. The ionization degree in gridded thrusters needs therefore only to be around 5--10\%.

\section{The physics of the PEGASES thruster}

Our team at Laboratoire de Physique des Plasmas (LPP) is developing a new thruster for space propulsion. This thruster is called PEGASES (for `plasma propulsion with electronegative gases') and belongs to the electric thrusters family. As for classical thrusters, the aim is to generate thrust and high specific impulse by the acceleration of charged particles. The innovative idea behind the PEGASES thruster is to generate thrust by both positive and negative ions, which has many advantages over existing technologies, such as eliminating the additional neutralization system and decreasing the weight and  size of the thruster.

\begin{figure}[htbp]
\centering\includegraphics[width=0.55\linewidth]{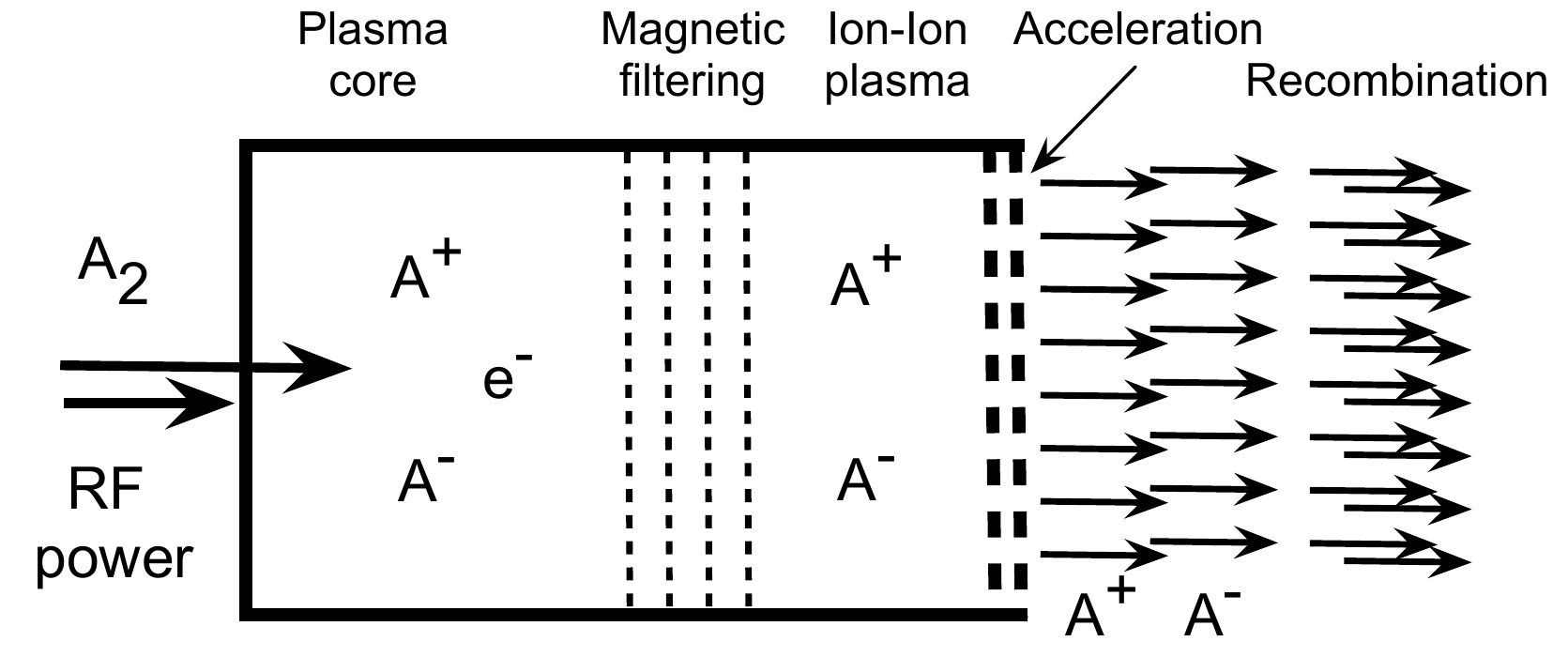}
\caption{Illustration of the PEGASES thruster, where positive and negative ions are accelerated alternately from an ion--ion plasma to generate thrust. }
\label{fig:pegases}
\end{figure}

The PEGASES concept is illustrated in Fig.\ \ref{fig:pegases} and can be described by three almost independent stages:

\begin{description}
\item[\bf{Stage 1}] is the plasma generation stage, where RF power is coupled to the plasma such that electrons are heated and cause ionization. This stage is important for the power efficiency of the thruster.

\item[\bf{Stage 2}] is the formation of an ion--ion plasma a certain distance from the ionization in Stage 1. For this we use a magnetic filter to confine and cool down the electrons. The hot electrons upstream of the barrier lead to high ionization and production of positive ions, while the cold electrons downstream of the barrier lead to efficient electron attachment and production of negative ions (for this to occur we need to use electronegative gases such as halogen-containing gases, e.g., O$_2$, SF$_6$ or I$_2$). As a result, positive and negative ions exist downstream with negligible contribution of electrons. This downstream region is described as an ion--ion plasma.

\item[\bf{Stage 3}] is the acceleration stage, where an alternate acceleration of positive and negative ions (from the ion--ion region in Stage 2) will provide the thrust for the spacecraft. The acceleration in PEGASES is based on classical gridded thrusters, where the acceleration is obtained by creating an electric field between two or more grids. In PEGASES the acceleration field is changed in time by applying a square waveform to the plasma grid; this allows consecutive bursts of positive and negative ion beams.
\end{description}

\subsection{Stage 1: plasma generation}

\begin{figure}[htbp]
\centering\includegraphics[width=0.9\linewidth]{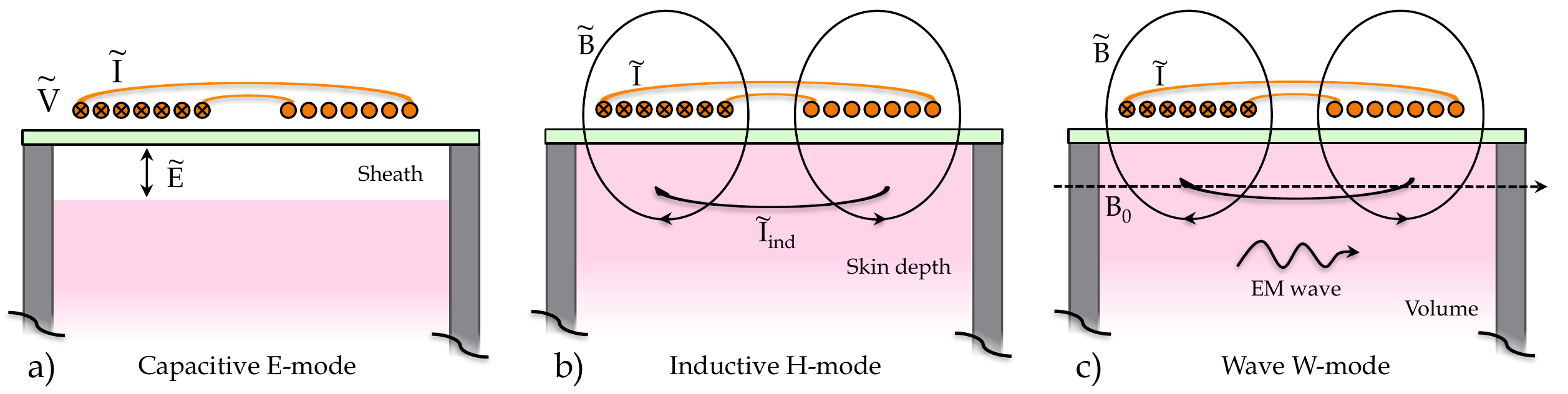}
\caption{Illustration of RF power discharges operated in the (a) capacitive, (b) inductive and (c) wave modes}
\label{fig:plasma-generation}
\end{figure}

RF plasmas are very often used in the semiconductor industry for etching and deposition, and a detailed description of the physics of these plasma discharges can be found elsewhere \cite{Chabert:2011book}. Briefly, the electromagnetic field from the RF power supply will couple to the electrons in the plasma and transfer energy to them. As illustrated in Fig.\ \ref{fig:plasma-generation}, there are a variety of methods to couple this energy to the electrons. For example, in Fig.\ \ref{fig:plasma-generation}(a) the RF voltage (from an electrode or a coil) couples capacitively to the electrons within an oscillating sheath. In Fig.\ \ref{fig:plasma-generation}(b) the RF current flowing in a coil induces an RF current in the plasma. The induced electric field transfers the energy to the electrons and consequently the field decays over a distance of a few centimetres in the plasma, called the skin depth. In Fig.\ \ref{fig:plasma-generation}(c) an additional static magnetic field allows an electromagnetic wave to propagate in the plasma and in this case transfers the energy to the electrons in the volume.

The plasma source in the PEGASES thruster is a purely inductively coupled plasma (ICP) without capacitive coupling, symmetrically driven at 4{\usp}MHz. The planar inductor is separated from the plasma by a thin (3{\usp}mm) ceramic window and encapsulated in a ferrite to reduce losses and to enhance the ICP coupling to the plasma \cite{godyak:2011,Aanesland:2012vs}. The RF power is fed to the inductor via an impedance matching network using a low-loss transmission-line transformer and air variable capacitors in symmetrical (push--pull) configuration. The symmetrical drive of the ICP inductor practically eliminates capacitive coupling to the plasma, resulting in negligible plasma RF potential. The power efficiency in such systems can reach up to 90\% or more \cite{godyak:2011} and is very important in space applications where every gram and every unit of electric power counts.

\subsubsection{Ionization and attachment in electronegative plasmas}

The plasma in the PEGASES thruster contains both positively and negatively charged ions.
To create negative ions, the gas has to be electronegative, allowing electrons to attach to  neutral species.
This process can occur in the volume or on surfaces with low work function.
In the negative ion sources for fusion, negative hydrogen ions are for example produced on a caesium surface \cite{Belchenko:1974tb,Bacal:2006iz}.

In most other experiments with negative ions, negative ion production occurs in volume by what we call dissociative attachment collisions.
Electrons are typically too energetic to attach directly to a neutral, so in the process either the excess energy leads to the neutral molecule breaking up and/or it dissipates into vibrational and rotational energies.
Electronegative gases are therefore molecular gases and typically formed by halogens (group 17 in the periodic table)
such as Cl$_2$ and I$_2$ or halogen-containing molecules such as SF$_6$ and CF$_4$. Although not halogens, O$_2$ and H$_2$ are also electronegative gases.
The ionization and attachment rate coefficients are typically increasing and decreasing functions of the  electron temperature, where the typical reactions are given in the form \cite{Rapp:1965bs}
\begin{eqnarray}
\mathrm{AB}_{x} + \mathrm{e} & \rightarrow & \mathrm{AB}_{x}^{+} + 2\mathrm{e}   \\
\mathrm{AB}_{x} + \mathrm{e} & \rightarrow & \mathrm{AB}_{y}^{-} + \mathrm{B}_z
\end{eqnarray}
for ionization and dissociative attachment. Hence, for most electronegative gases, ionization dominates for high electron temperatures whereas attachment occurs for low temperatures.

\subsection{Stage 2: ion--ion plasma formation}

\begin{figure}[htbp]
\centering\includegraphics[width=0.98\linewidth]{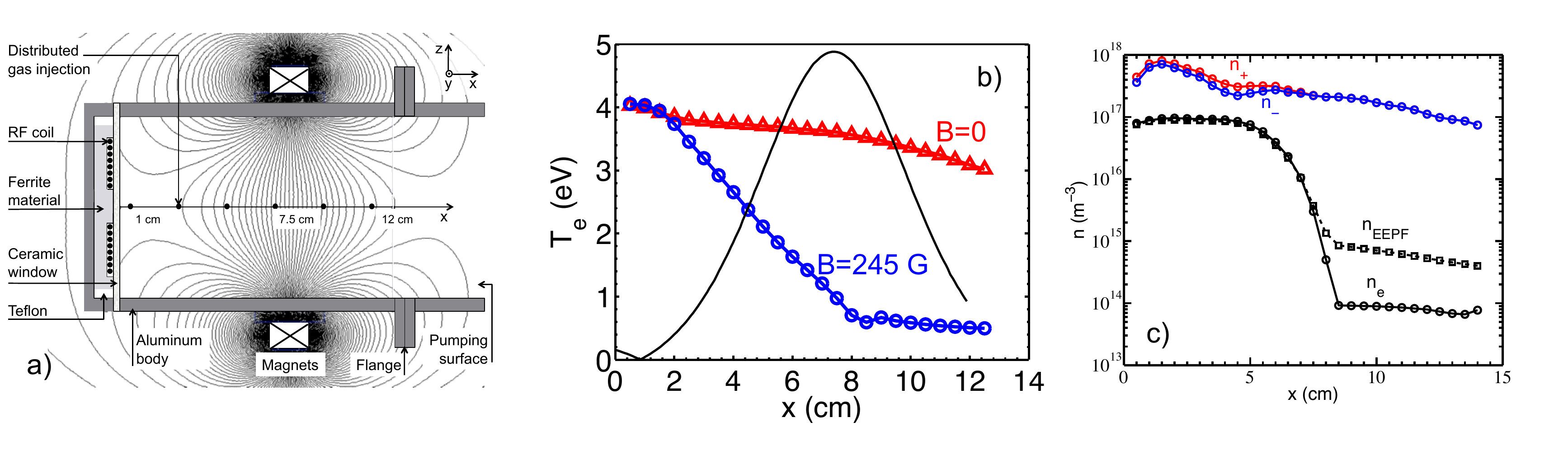}
\caption{Illustration of RF power discharges operated in the (a) capacitive, (b) inductive and (c) wave modes}
\label{fig:stage2-filter}
\end{figure}

The aim for this stage is to segregate the plasma into two regions: (i)~the plasma core, where electrons have a rather high temperature for efficient ionization; and (ii)~a downstream ion--ion plasma, where the electron density can be neglected and the plasma dynamics is controlled by ions.
As seen above, to acquire this segregation, we need to control the electron temperature in the plasma.
Commonly, in low-temperature plasmas, the electron temperature $T_\rme$ is governed by the ionization balance (electron creation and loss processes) and is a function of the kind of gas and the product $pL$, where $p$ is the gas pressure and $L$ is the characteristic size of the plasma \cite{lieberman:2005book}.
The specific mechanism of electron heating seen in Fig.\
\ref{fig:plasma-generation} and the value of discharge power have a minor influence on the electron temperature.
In negative ion sources, the electron cooling is usually achieved with magnetic filters (a localized transverse magnetic field) placed in front of the ion extracting aperture. This technique is also used in the PEGASES thruster, where a localized magnetic barrier is generated by a set of permanent neodymium magnets forming a localized Gaussian magnetic field perpendicular to the extraction axis. The illustration of the PEGASES prototype with the magnetic field lines is shown in Fig.\ \ref{fig:stage2-filter}(a).
We have recently shown that the gradient and strength of the magnetic field strongly affect the electron cooling, and the position of the minimum electron temperature is achieved at the maximum magnetic field strength \cite{Aanesland:2012vs}. We have also shown that the ion--ion plasma is also generated around the maximum field region \cite{Bredin:2012wg}. As an example, the electron temperature measured in argon, and the measured densities in SF$_6$ are shown in Fig.\ \ref{fig:stage2-filter}.
Close to the maximum magnetic field, the electron density is three orders of magnitude lower than the ion densities. The ion densities in this region remain high, showing that ion transport across the magnetic field is not affected by the magnetic field. For an RF power of only 120{\usp}W, the ion density reaches $3\times10^{17}${\usp}m$^{-3}$ in the ion--ion region.

\subsection{Stage 3: alternate acceleration of ions to generate thrust}

\begin{figure}[htbp]
\centering\includegraphics[width=0.5\linewidth]{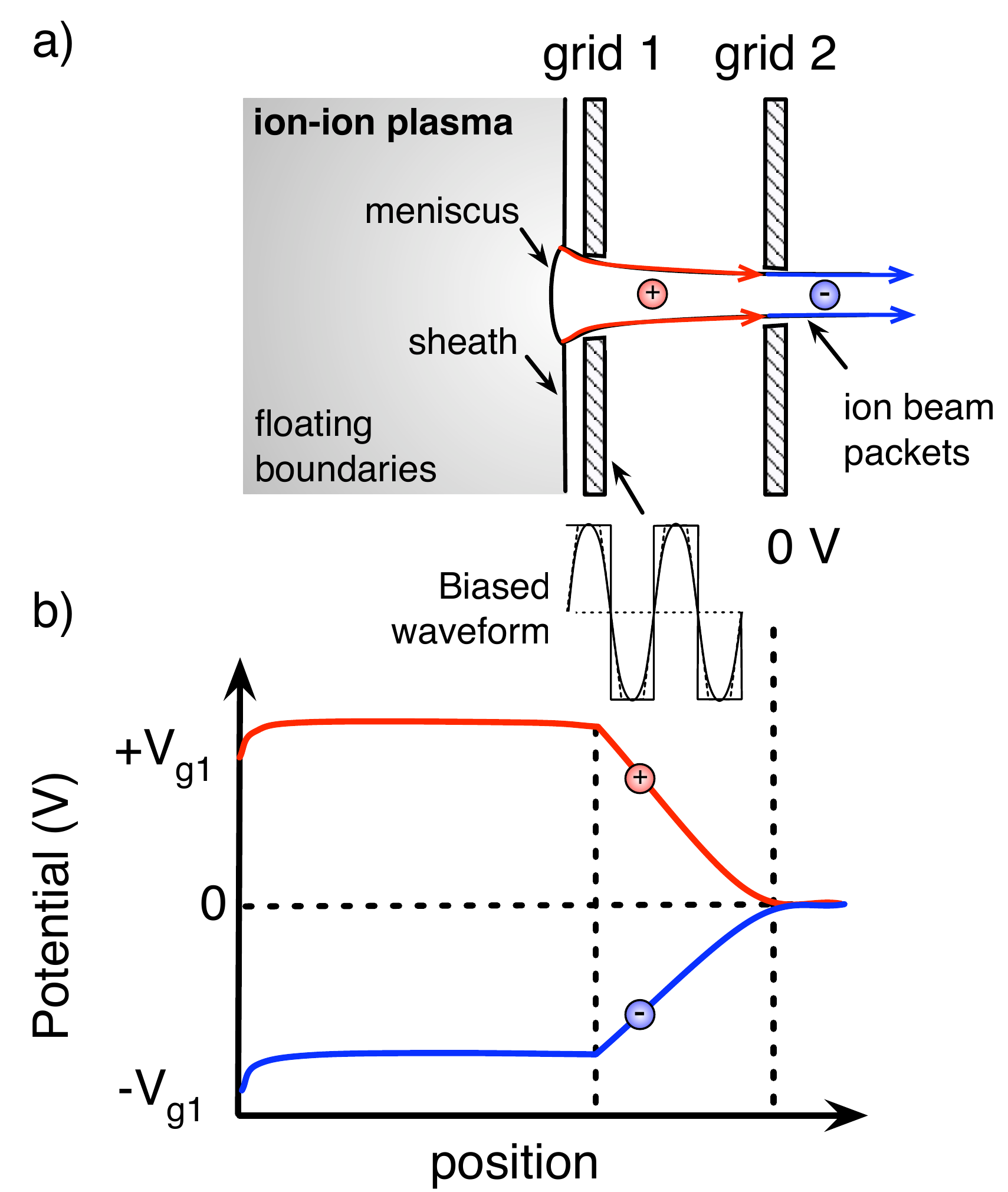}
\caption{Illustration of RF power discharges operated in the (a) capacitive, (b) inductive and (c) wave modes}
\label{fig:Alternate-acceleration-scheme}
\end{figure}

The PEGASES thruster is a gridded thruster based on the same principles as classical gridded acceleration systems.
However, the originality of this thruster is that positive and negative ions are alternately accelerated from an ion--ion plasma using the same grids.
To achieve this, the first grid in contact with the plasma is biased with alternate voltage waveforms and the second grid is grounded \cite{Aanesland:2012uz}.
The walls in the thruster are floating such that the potential of the plasma follows the plasma grid potential, and hence the electric field between the two grids will change direction during one bias period.
A simplified illustration of this concept is shown in Fig.\ \ref{fig:Alternate-acceleration-scheme}, where (a) shows one hole in the grid system with idealized ion trajectories, and (b) shows the potential distribution in the plasma and across the grids for the positive and negative bias period applied to the first grid. Note that possible variations of the plasma potential within the bulk plasma are not shown here.

The acceleration grids in any ion source are designed and adapted to the acceleration voltage and the available plasma parameters (density and temperature), to achieve the desired beam properties with respect to thrust, beam current, divergence, etc. Square waveforms are therefore best suited. Particle-in-cell simulations have shown that the ratio of extracted positive and negative ion fluxes from the plasma depends strongly on the electronegativity, particularly when the negative ion density $n_{\rmi -}$ is much higher than the electron density $n_{\rme}$  \cite{Oudini:2013uy}.
When negative ions are extracted from the plasma, we should therefore expect co-extracted electrons. In this case, the period over which the negative charges are accelerated should be shorter than the corresponding period for positive ions in order to compensate both the charge density and the current (i.e., the duty cycle can be optimized). This asymmetry could even be pushed towards a system of accelerating positive ions and electrons. For an optimized system, the acceleration voltages for the positive and negative charges should also be adapted independently to compensate for the difference in their effective temperatures.
The most suitable waveform is therefore a square waveform with adjustable duty cycle and voltage offset.

The upper and lower frequency requirements for the applied voltage should take into account: (i)~both the ion plasma frequency (so the ions can react to the variations) and the transit time of the ions through the grids; and (ii)~the potential barrier created in the downstream space from the space charge of the single beam packet or envelope, respectively. Analytical models developed in one dimension, which completely neglect any secondary electron emission or other external sources of space-charge  neutralization, predict an operating frequency in the lower megahertz region \cite{aanesland:HDR}.

\subsection{First laboratory tests in PEGASES II}

\begin{figure}[htbp]
\begin{center}
\includegraphics[width=1.0\textwidth]{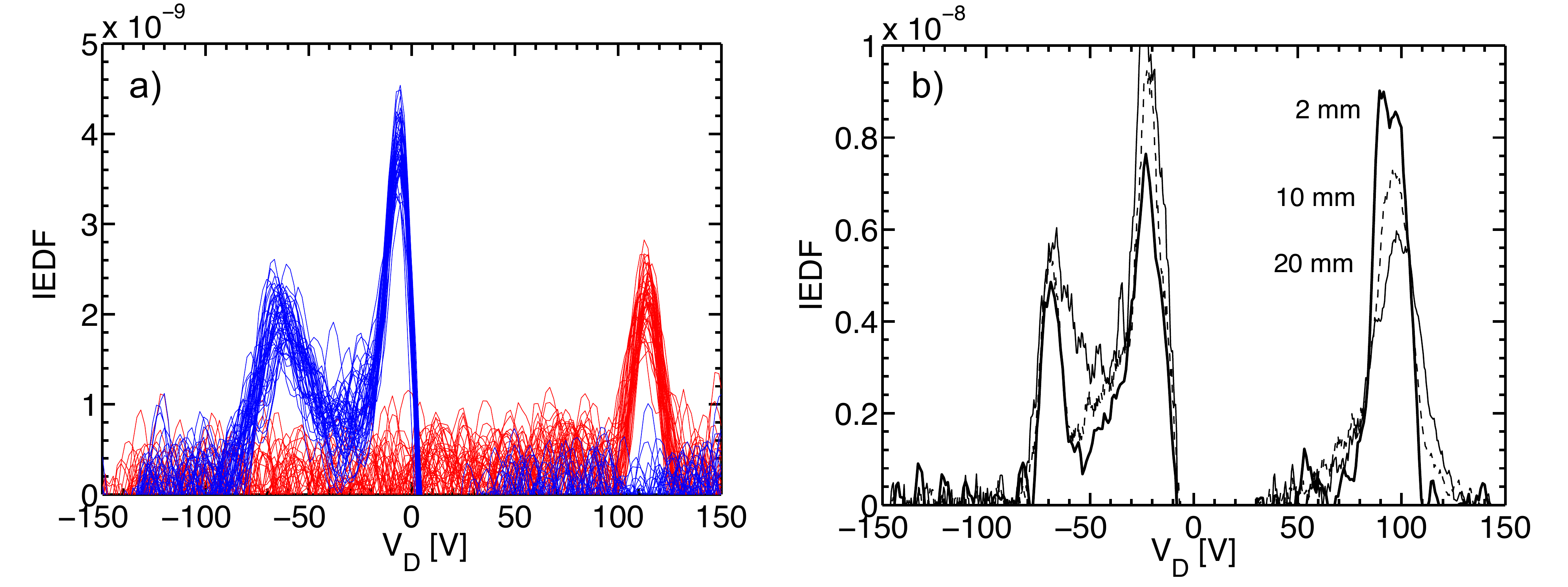}
\caption{(a) IEDFs measured during the bias period, where the red and blue curves are measured during the positive and negative bias period, respectively. (b)~IEDF measured continuously for various positions downstream of the grounded grid. The acceleration voltage is here $V_0=\pm 100${\usp}V at a bias frequency of 1{\usp}kHz. }
\label{fig:alternte-acceleration-results-1}
\end{center}
\end{figure}

Successive positive and negative ion beams have been measured downstream of the alternately biased grids in the PEGASES~II thruster. The grids are placed in the ion--ion plasma region and the bias frequency is only 1{\usp}kHz biased.
Figure\ \ref{fig:alternte-acceleration-results-1}(a) shows a typical result of time-resolved ion energy distribution functions (IEDFs) measured downstream of the acceleration stage. The red and blue curves are obtained during the positive and negative bias periods, respectively.
Both the positive and negative ions are measured with a constant amplitude and energy during their respective bias period. However, their mean energies are not the same. Figure\ \ref{fig:alternte-acceleration-results-1}(b) shows the IEDFs measured for various positions downstream of the grounded grid. These measurements are continuous while the grid is alternately biased. The peak ion energies remain constant downstream of the grid, but interestingly the positive peak amplitude decreases while the negative ion peak increases as a function of position.

\begin{figure}[htbp]
\begin{center}
\includegraphics[width=0.9\textwidth]{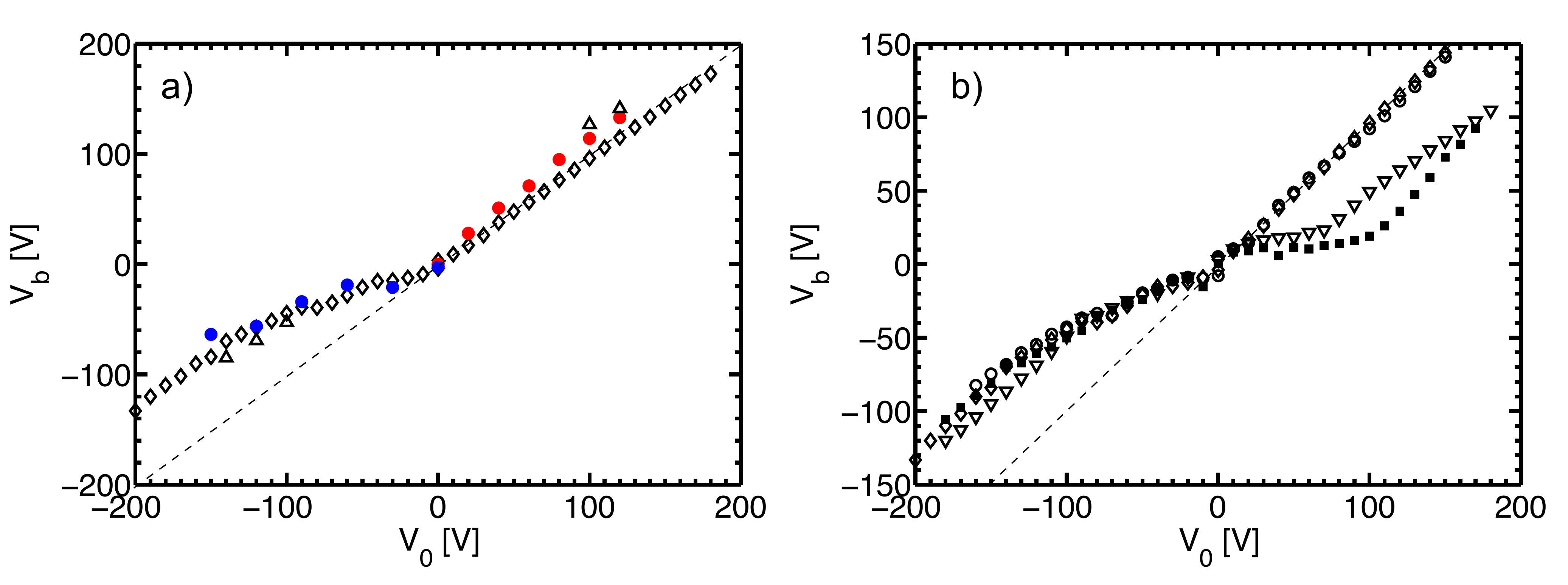}
\caption{Ion mean energy as a function of the acceleration voltage. (a) Circles and diamonds are obtained for alternate and continuous acceleration, respectively, and measured with the Retarding Field Energy Analyser (RFEA). Triangles are measured with a planar probe in continuous mode. (b)~Increasing deposition of a fluorine--sulfur film on the acceleration grid. Diamonds are the initial conditions, triangles after about 5{\usp}min and diamonds after 30{\usp}min of operation, and circles after cleaning the grids.}
\label{fig:alternte-acceleration-results-2}
\end{center}
\end{figure}

Figure\ \ref{fig:alternte-acceleration-results-2}(a) shows the ion mean energies as a function of the applied acceleration voltage for continuous and alternate acceleration operated in a mixture of argon and SF$_6$ (about 20\% SF$_6$).
The positive ions have energies equivalent to the acceleration voltage $V_0$. Some variations are obtained between each dataset, possibly due to slight differences in the ion--ion plasma conditions and therefore a change in the sheath voltage in front of the grids.
The negative ions are, on the contrary, measured with lower energies than expected from the applied acceleration voltage. Similar results are obtained with planar probes. Figure\ \ref{fig:alternte-acceleration-results-1}(b) shows that the beam energy does not depend on the downstream position, indicating that the bipolar beam is space-charge-neutralized downstream.

The difference in the positive and negative ion energies can be understood as follows.
In plasma immersed ion implantation (PIII) sudden negative bias pulses are applied to a surface such that positive ions are implanted in the material  \cite{Lieberman:1989vh}. Several authors have seen that implanting into a dielectric substrate results in a significant voltage build-up in the wafer, reducing the effective implant energy \cite{Cheung:1996tm,Fu:2004gy}.
In our case, a film composed of fluorine and sulfur can deposit on the acceleration grids. A dielectric film is observed by eye on the grids after some hours of operation in mixed Ar/SF$_6$ and after some tens of minutes in pure SF$_6$. This film is also detectable by measuring a decrease in the current on the grids.
In the presence of a dielectric layer on the plasma grid, a certain voltage will be dropped in the dielectric such that the effective acceleration voltage is \cite{Fu:2004gy}
\begin{equation}
V_{\mathrm{eff}} = V_0 - \frac{e\delta Q}{\varepsilon_0\varepsilon_{\rm r}},
\end{equation}
where $\delta$ is the thickness of the dielectric film and $Q$ is the surface charge.
For a quick estimate using $\delta \sim 1{\usp}\rmmu$m, $Q \sim 1\times10^{16}$ and $\varepsilon_{\rm r} = 3.7$ as for pure sulfur, the voltage drop in the dielectric is about 50{\usp}V, which is of the order of the measured energy loss for negative ions.
Figure\ \ref{fig:alternte-acceleration-results-2}(b) shows the beam energies as a function of $V_0$ in pure SF$_6$ after about 5{\usp}min and 30{\usp}min of operation, i.e., the deposit on the grids became thicker and thicker.  The negative ions are not affected much with increasing film thickness. However, the positive ion energies are similar to the negative ion energies after some time of operation.
It seems likely that it is the dielectric film that is responsible for the lower beam energies. However, it is still uncertain why the negative ions are affected more than the positive ions in the initial conditions.
This might be due to a change in the dielectric layer formed as a function of the bias voltage or a change in how this dielectric is charged when applying a positive or negative bias.
Further experiments are needed before any definite conclusion can be drawn.

\section{Conclusion}

The PEGASES thruster is a new and promising gridded ion thruster where, in contrast to classical gridded thrusters, both positive and negative ions are accelerated alternately from the same source. This concept makes the additional neutralization system redundant and provides therefore many advantages over existing thrusters.
The summary given here provides a brief overview of the state of the art in the development of this thruster. Although there is still a long way to go before this thruster is flying in space, we show that an ion--ion plasma is efficiently produced downstream of a localized magnetic filter, such that the degree of ionization required in a gridded thruster is met. The ion--ion plasma source can therefore provide high enough densities for gridded ion thrusters operating with an acceleration of 500{\usp}V or so, and can therefore provide similar thrust and $I_{\rm sp}$ as classical gridded thrusters.
We have also shown the first laboratory evidence that alternate acceleration of positive and negative ions can be achieved.

\section{Acknowledgements}

This work was funded by EADS Astrium and by Agence Nationale de la Recherche (ANR) under contract ANR-11-BS09-040.
The authors would like to thank V.~Godyak, J.-P.~Booth and G.~Hagelaar for useful discussions and collaboration.


\begin{thebibliography}{99}

\bibitem{kaufman:1985aj0} 
H.R. Kaufman, Technology of closed-drift thrusters. {\em AIAA J.} \textbf{23} (1985) 78--87.

\bibitem{goebel:20080} 
D.M. Goebel and I. Katz, {\em Fundamentals of Electric Propulsion: Ion and Hall Thrusters} (JPL Space Science and Technology Series, Wiley, Hoboken, NJ, 2008).

\bibitem{hayabusa} 
N. Kazutaka and H. Kuninaka, {\em Trans. Jpn. Soc. Aeronaut. Space Sci., Aerosp. Technol. Jpn.} \textbf{10} (2012) Tb-1.

\bibitem{Aanesland:2009jm} 
A. Aanesland, A. Meige, and P. Chabert, Electric propulsion using ion--ion plasmas. {\em J. Phys.: Conf. Ser.} \textbf{162} (2009) 012009.
    
\bibitem{Janovsky}    
R. Janovsky and D.E. Koelle, Development and transportation costs of space launch systems. Presented at the DGLR/CEAS European Air and Space Conference (2007).

\bibitem{kaufman:1982} 
H.R. Kaufman, Technology and applications of broad-beam ion sources used in sputtering. Part I. Ion source technology. {\em J. Vac. Sci. Technol.} \textbf{21} (1982) 725.

\bibitem{Chabert:2012dh} 
P. Chabert, J. Arancibia Monreal, J. Bredin, L. Popelier, and A. Aanesland, Global model of a gridded-ion thruster powered by a radiofrequency inductive coil. {\em Phys. Plasmas} \textbf{19}(7) (2012) 073512.

\bibitem{Chabert:2011book} 
P. Chabert and N.S.J. Braithwaite, \textit{Physics of Radio-Frequency Plasmas} (Cambridge University Press, Cambridge, 2011).

\bibitem{godyak:2011} 
V.A. Godyak, Electrical and plasma parameters of ICP with high coupling efficiency. {\em Plasma Sources Sci. Technol.} \textbf{20} (2011) 025004.

\bibitem{Aanesland:2012vs} 
A. Aanesland, J. Bredin, P. Chabert, and V.A. Godyak, Electron energy distribution function and plasma parameters across magnetic filters. {\em Appl. Phys. Lett.} \textbf{100} (2012) 044102.

\bibitem{Belchenko:1974tb} 
Yu.I. Belchenko, G.I. Dimov, and V.G. Dudnikov, A powerful injector of neutrals with a surface-plasma source of negative ions. \textit{Nucl. Fusion} \textbf{14}(1) (1974) 113.

\bibitem{Bacal:2006iz} 
M. Bacal, Physics aspects of negative ion sources. {\em Nucl. Fusion} \textbf{46} (2006) S250--S259.

\bibitem{Rapp:1965bs} 
D. Rapp and D.D. Briglia, Total cross sections for ionization and attachment in gases by electron impact. II. Negative-ion formation. {\em J. Chem. Phys.} \textbf{43} (1965) 1480.

\bibitem{lieberman:2005book} 
Lieberman, M. A., and Lichtenberg, A. J., Principles of Plasma Discharges and Materials Processing. (2nd ed.). Wiley-Interscience (2005).

\bibitem{Bredin:2012wg} 
J. Bredin, P. Chabert, and A. Aanesland, Langmuir probe analysis of highly electronegative plasmas. {\em Appl. Phys. Lett.} \textbf{102} (2013) 154107.

\bibitem{Aanesland:2012uz} 
A. Aanesland, P. Chabert, M. Irzyk, and S. Mazouffre, Method and device for forming a plasma beam. Patent WO 2012/042143 A1 (2012).

\bibitem{Oudini:2013uy} 
N. Oudini, J.-L. Raimbault, P. Chabert, A. Meige, and A. Aanesland, Particle-in-cell simulation of an electronegative plasma under direct current bias studied in a large range of electronegativity. {\em Phys. Plasmas} \textbf{20} (2013) 043501.

\bibitem{aanesland:HDR} 
A. Aanesland, Low temperature plasmas exposed to external magnetic and electric fields. Habilitation \`a Diriger des Recherches, Universit\'e Pierre et Marie Curie, 2013.

\bibitem{Lieberman:1989vh} 
M.A. Lieberman, Model of plasma immersion ion implantation. {\em J. Appl. Phys.} \textbf{66} (1989) 2926.

\bibitem{Cheung:1996tm} 
N. Cheung, Plasma immersion ion implantation for semiconductor processing. \textit{Mater. Chem. Phys.} \textbf{46} (1996) 132--139.

\bibitem{Fu:2004gy} 
R.K.Y. Fu, Influence of thickness and dielectric properties on implantation efficacy in plasma immersion ion implantation of insulators. \textit{J. Appl. Phys.} \textbf{95}(7)  (2004) 3319.

\end{thebibliography}
\end{document}